\documentclass[12pt]{article}

\usepackage{graphicx}
\usepackage{bm}
\usepackage{amsmath}
\usepackage{amssymb}

\begin{document}

\title{Bounded fitness landscapes and the evolution of the linguistic diversity}
\author{Viviane M. de Oliveira$^{a}$\thanks{viviane@df.ufpe.br}, Paulo R. A. Campos$^{b}$, M. A. F. Gomes$^{a}$, \\ I. R. Tsang$^{c}$}

\maketitle

\noindent
$^{a}$Departamento de F\'{\i}sica, Universidade Federal de Pernambuco, 50670-901, Recife, PE, Brazil\\
\noindent
$^{b}$Departamento de F\'{\i}sica e Matem\'atica,  Universidade Federal
Rural de Pernambuco 52171-900, Dois Irm\~aos, Recife-PE, Brazil\\
\noindent
$^{c}$Centro de Inform\'atica, Universidade Federal de Pernambuco, 50670-901, Recife, PE, Brazil

\bigskip

\begin{abstract}
A simple spatial computer simulation model was recently introduced to study the evolution of the linguistic diversity \cite{Viviane}. The model considers ~ processes of selective geographic colonization, linguistic anomalous diffusion and mutation. In the approach, we ascribe to each language a fitness function which depends on the number of people that speak that language. Here we extend the aforementioned model to examine the role of saturation of the fitness on the language dynamics. We found that the dependence of the linguistic diversity on the area after colonization displays a power law regime with a nontrivial exponent in very good agreement with the measured exponent associated with the actual distribution of languages on the Earth.
\end{abstract}

\section{Introduction}

The research in language dynamics has arose an increasing interest of the complex systems community in the last years.  Most of the researchers focus their investigations on issues like rise, competition, extinction risk and death of languages \cite{Sutherland, Abrams, Patriarca, Schulze, Mira, Schwammle, Kosmidis, Stauffer2, Tesileanu, Diamond}. Furthermore, recent advances in archeology, genetics and linguistics have provided relevant contributions to a better comprehension of the linguistic diversification \cite{Renfrew2, Cavalli-Sforza}. Some investigations have demonstrated that distinct causes have greatly affected the evolution of the linguistic diversity. Among the main elements are geographic factors, economic features, complexity of the language, to cite just a few. For instance, Sutherland \cite{Sutherland} has shown that beside country area, forest area and maximum altitude contribute to increase diversity, whereas the diversity decreases for a larger latitude. According to Bellwood \cite{Bellwood, Bellwood2} and Renfrew \cite{Renfrew, Renfrew3} the occurrence of agricultural expansion was the responsible for the massive population replacements initiated about 10,000 years ago and caused the disappearance of many of the Old World languages.

In a recent work, we investigated the evolution of the linguistic diversity by introducing a spatial computer simulation model that considers a diffusive process which is able to generate and sustain the diversity \cite{Viviane}. The model describes the occupation of a given area by populations speaking several languages. To each language was assigned a fitness value $f$ which is proportional to the number of sites colonized by populations that speak that language. In the process of colonization, language mutation or differentiation and language substitution can take place, which affords the linguistic diversity. This simple model gives rise to scaling laws in close resemblance with those reported in \cite{Gomes1}.

In the current contribution, we study the dynamics of the linguistic diversity but now we assume that the fitness of each language is bounded by a given maximum (saturation) value which is randomly chosen from an uniform distribution. The saturation hypothesis mimics factors like the difficulty/ease of learning the languages and economy that permit some languages to propagate more easily than others.

The paper is organized as follows. In Section 2 we introduce the model. In Section 3 we discuss the results. And finally, in Section 4 we present the conclusions.

\section{Model}

Our model is defined on a two-dimensional lattice of linear size $L$, and composed of $A=L \times L$ sites with periodic boundary conditions. Each lattice site $s_i$ represents a given region, which can be occupied by a single population speaking just one language. We ascribe to each site a given capability $C_i$, whose value we estimate from a uniform distribution, defined in the interval 0-1. The capability means the amount of resources available to the population which will colonize that place. It is implicit that the population size in each cell $s_i$ is proportional to its capability $C_i$.

In the first step of the dynamics, we randomly choose one site of the lattice to be colonized by a single population that speaks the ancestor language. Each language is labeled by an integer number. As soon as a new language arises, it is labeled by the next upper integer. To each language, we assign a fitness value $f$, which is calculated as the sum of the capabilities of the sites which speak that specific language. But now differently from reference \cite{Viviane}, the fitness can not exceed an integer value $\gamma_k$ which we have chosen to be in the range 1-2000. This saturation term $\gamma_k$ is randomly chosen when the language $k$ appears. Thus, the initial fitness of the ancestor language is the capability of the initial site.

In the second step, one of the four nearest neighbors of the site containing the ancestor language will be chosen to be colonized with probability proportional to its capability. We assume that regions containing larger amount of resources are most likely to be colonized faster than poor regions. The referred site is then occupied by a population speaking the ancestor language or a mutant version of it. Mutations are the mechanisms responsible for generating diversity, and together with the natural selection mantains the standing level of diversity on the system. The probability of occurrence of a mutation in the process of propagation is $p=\frac{\alpha}{f}$, where $\alpha$ is a constant, and so the mutation probability is inversely proportional to the fitness of the language. The form of the mutation probability $p$ is inspired by population genetics, where the most adapted organisms are less likely to mutate than poorly adapted organisms \cite{Barton}. The probability of producing reverse mutations is zero, that is, the language generated by a mutation is always different of the previous ones.

\begin{figure}
\centering
\includegraphics[height=11cm,width=11cm,angle=0]{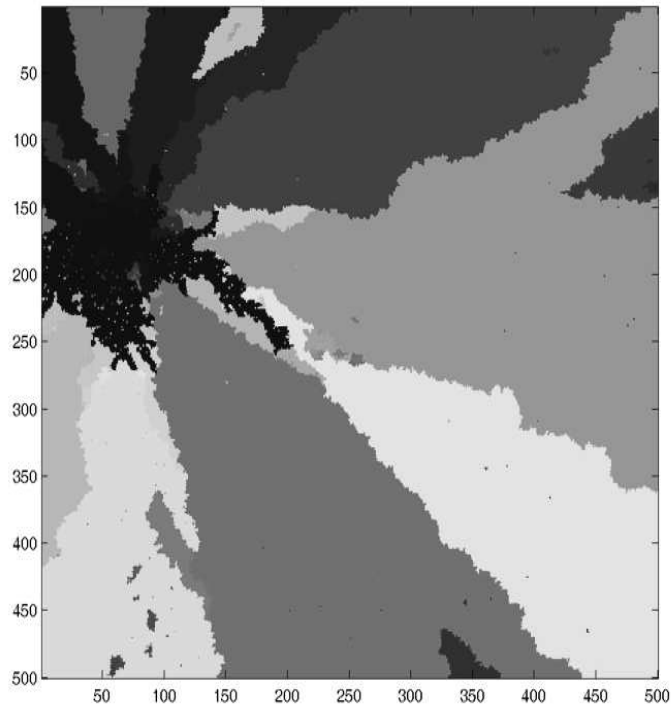}
\caption{Snapshot of a typical realization of the dynamics at the first moment of colonization of all sites. The saturation quantities $\gamma_k$ are randomly chosen in the interval 1-2000. The lattice size is $L=500$ and $\alpha=0.3.$ See text for detail.}
\end{figure}

In the subsequent steps, we check the empty sites which are located on the boundary of the colonized cluster, and we then choose one of those empty sites according to their capabilities. Again, those sites with higher capabilities enjoy of a greater likelihood to be occupied. After that, we choose the language to be incorporated in the chosen cell among those languages occupying the neighboring sites. Languages with higher fitness have higher chance to expand. The process continues while there are empty sites in the network. After completion, we count the total number of languages $D$. In order to give to the reader some insight about our model, in Figure 1 we present the snapshot for a typical realization of the dynamics at the first moment of colonization of all sites (in this figure the gray scale represents different languages). The striated linguistic domains presenting very small territories occupied by different languages shown in Figure 1 remind us the actual distribution of languages observed in the Caucasus region between Black and Caspian Seas, a relatively small area of 300,000 km$^2$ where languages of the Caucasic, Indo-European and Altaic families coexist distributed within a large variety of peoples \cite{Cavalli-Sforza}.

\section{Results and Discussion}

In Figure 2, we show the diversity $D$ as a function of the area $A$ (total number of sites in the lattice) for mutation parameter $\alpha=0.3$ and saturation values defined in the interval 1-2000. The points are averages over 100 independent simulations when $L<400$ and over 20 simulations when $L=500$. We observe that the curve presents just one scaling region which extends over five decades. The exponent $z=0.39 \pm 0.01$ is in quite satisfactory agreement with the exponent observed for the actual distribution of languages on Earth. For sake of completeness, we also exhibit in Figure 2 the observed values ($\ast$) of diversity versus area obtained in reference \cite{Gomes1} for all languages spoken on Earth (the ten data points are associated with the interval from $A=50$ km$^2$ to $A=10^7$ km$^2$ of the actual distribution). We notice in passing that although there is not a perfect scaling relationship between diversity and area along five decades in area, both the simulation and the actual data of $D(A)$ curiously seem to be modulated by a similar tendency to oscillate in respect to the main scaling behavior (the deviations from perfect scaling in the actual data have no connection with the choice of the bins). We have also investigated the situation at which the saturation value is the same for all languages. We have noticed a linear growth of the diversity with area when the maximum $\gamma$ is very small. For large values of $\gamma$ we notice the existence of two scaling regions. For very large values of $\gamma$ we recover the result obtained for the case where the fitness are not limited \cite{Viviane}.

\begin{figure}
\centering
\includegraphics[height=12cm,width=12cm,angle=270]{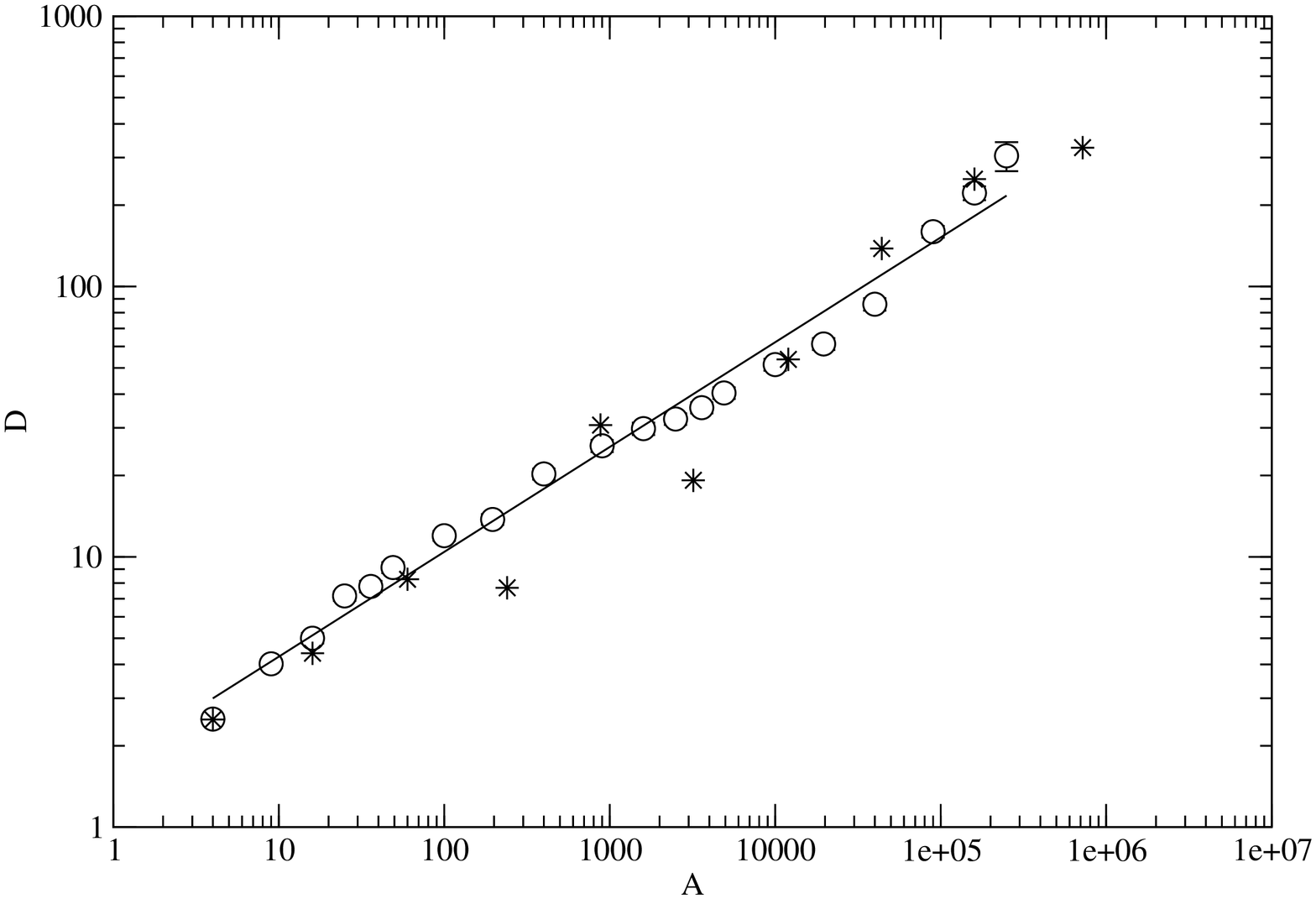}
\caption{ Number of languages $D$ as a function of the area $A$ for $\alpha=0.3$. The exponent is $z=0.39 \pm 0.01$. The asterisks represent data from the actual distribution of languages on Earth. See text and Figure 1 of reference \cite{Gomes1} for detail.}
\end{figure}

\begin{figure}
\centering
\vspace{-0.5cm}
\includegraphics[height=12cm,width=12cm,angle=270]{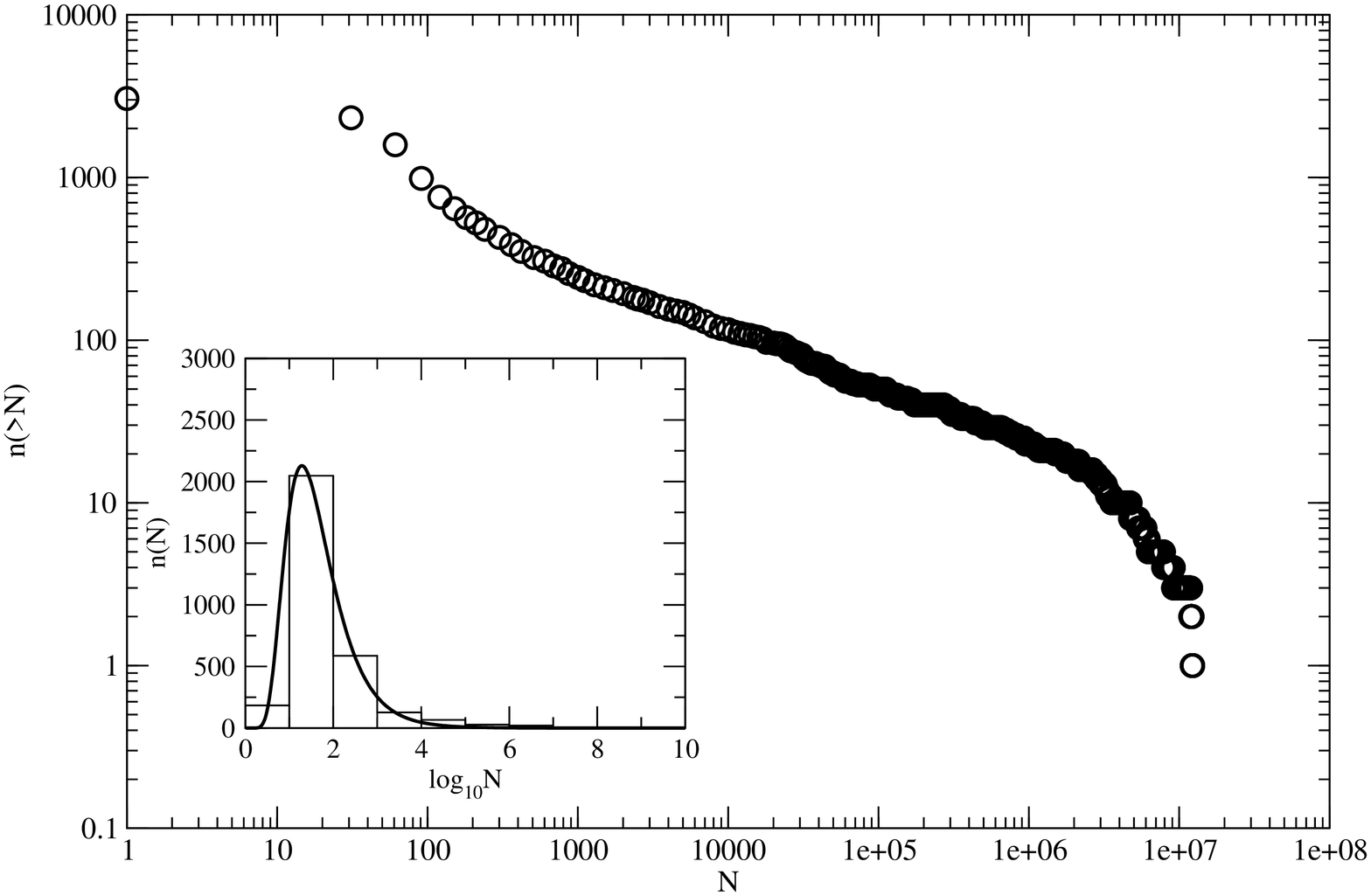}
\caption{\label{fig:figure3} Main plot - number of languages with population greater than $N$, $n(>N)$, as a function of $N$. $n(>N) \sim N^{-\tau}$ with $\tau=0.35 \pm 0.01$ for $200<N<2,000,000$ and $\tau=1.14 \pm 0.01$ for $2,000,000<N<10,000,000$. Inset - corresponding differential distribution $n(N)$ with lognormal best fit (continuous line). See text for detail.}
\end{figure}

Figure 3 displays the number of languages with population size greater than $N$, $n(>N)$, as a function of $N$. In order to obtain the curves, we have assumed that the population in a given site is proportional to the capability in the site. We have considered that the population in a given site is its capability multiplied by a factor 100. In the plot, the values of the parameters are $L=500$ and $\alpha=0.3$. In close analogy with the distribution of languages on Earth \cite{Gomes1}, we find two distinct scaling regimes $n(>N) \sim N^{-\tau}$: $\tau=0.35 \pm 0.01$ for $200<N<2,000,000$, and $\tau=1.14 \pm 0.01$ for $2,000,000<N<10,000,000$. The inset exhibits the differential distribution of languages  spoken by a population of size $N$, $n(N)$. This distribution also agrees with the one observed for languages on Earth \cite{Gomes1, Sutherland}; in particular it is well described by the lognormal function $n(N)=\frac{1}{\sqrt{2\pi}\sigma}\frac{1}{N} \exp \left[-\frac{1}{2\sigma^2}(\log N- \mu)^2 \right]$, with $\sigma=0.41$ and $\mu=0.42$ (continuous curve in the inset).

\section{Conclusions}

We have introduced a model for evolution of linguistic diversity that considers a bounded fitness value for languages. We have considered a random chosen value of saturation of the fitness for each language in order to mimic the fact that different languages have different conditions to propagate. We have noticed a considerable improvement of the results when compared to the earlier approach \cite{Viviane}. Now, the relationship between diversity and area presents just one scaling regime. For $\alpha=0.3$ we obtain $z=0.39\pm 0.01$, which is in very good agreement with the exponent observed for the languages on the Earth \cite{Gomes1}, along five decades of variability in area. We have also observed that the exponents $\tau$ for the two power law regimes in $n(>N)$ as a function of $N$ are closer to those obtained by empirical observations \cite{Gomes1}.

In order to compare other kinds of saturation conditions, we have also studied the case where the saturation values are the same for all the languages. With this condition, we could not reproduce the basic relationship between diversity and area observed for the actual distribution of languages, although for the very particular and unrealistic case where $\alpha=0.01$ and $\gamma=1$, we can perfectly reproduce the differential distribution of languages spoken by a population of size $N$, $n(N)$, as well as the number of languages with population size greater than $N$, $n(>N)$, as a function of $N$. Our results seem to demonstrate that different assumptions on the behavior of the fitness function have very important consequences on the characteristics of the language spreading.

V.~M. de Oliveira and M.~A.~F. Gomes are supported by Conselho Nacional de Desenvolvimento Cient\'{\i}fico e Tecnol\'ogico and Programa de N\'ucleos de Excel\^encia (Brazilian Agencies). P.~R.~A. Campos is supported by CNPq.

\end{document}